\documentclass[%
 reprint,
 groupedaddress,
 amsmath,amssymb,
 aps,prl,
 floatfix,showkeys
]{revtex4-1}

\usepackage[utf8]{inputenc}
\usepackage{lmodern}  
\DeclareUnicodeCharacter{2212}{-}

\usepackage[english]{babel}
\makeatletter
\def\bbl@set@language#1{%
  \edef\languagename{%
    \ifnum\escapechar=\expandafter`\string#1\@empty
    \else\string#1\@empty\fi}%
  \@ifundefined{babel@language@alias@\languagename}{}{%
    \edef\languagename{\@nameuse{babel@language@alias@\languagename}}%
  }%
  \select@language{\languagename}%
  \expandafter\ifx\csname date\languagename\endcsname\relax\else
    \if@filesw
      \protected@write\@auxout{}{\string\select@language{\languagename}}%
      \bbl@for\bbl@tempa\BabelContentsFiles{%
        \addtocontents{\bbl@tempa}{\xstring\select@language{\languagename}}}%
      \bbl@usehooks{write}{}%
    \fi
  \fi}
\newcommand{\DeclareLanguageAlias}[2]{%
  \global\@namedef{babel@language@alias@#1}{#2}%
}
\makeatother
\DeclareLanguageAlias{en}{english}
\addto\captionsenglish{}

\usepackage{graphicx}
\usepackage{xfrac}
\usepackage{xspace}
\usepackage{siunitx}
\usepackage{dcolumn}
\usepackage{natbib}
\usepackage{hyperref}
\usepackage[dvipsnames]{xcolor}

\usepackage[capitalize]{cleveref}
  \crefformat{equation}{Eq.~(#2#1#3)}
  \crefrangeformat{equation}{Eqs.~(#3#1#4) to~(#5#2#6)}
  \crefformat{figure}{Fig.~#2#1#3}
  \crefrangeformat{figure}{Figs.~#3#1#4 to~#5#2#6}

\let\baraccent=\= 
\newcommand{\emath}{\ensuremath}

\DeclareRobustCommand{\kt}{\emath{k{\!\!\;\scriptstyle{_B}}T}\xspace}
\DeclareRobustCommand{\Dt}{\Delta t\xspace}

\DeclareSIUnit[number-unit-product = {\,}]\cal{cal}
\DeclareSIUnit\kcal{\kilo\cal}
\DeclareSIUnit\kJ{\kilo\joule}

\renewcommand{\=}[1]{\stackrel{#1}{=}} 

\DeclareRobustCommand{\pd}[1]{\partial #1}
\DeclareRobustCommand{\pdf}[2]{\frac{\pd #1}{\pd #2}}
\DeclareRobustCommand{\pdi}[2]{\pd_{#2} #1}

\DeclareRobustCommand{\ii}[2]{\int_{#1}^{#2}}

\DeclareRobustCommand{\avg}[1]{\left\langle #1 \right\rangle} 

\newcommand{\abs}[1]{\left| #1 \right|} 

\DeclareRobustCommand{\xbar}[1]{%
   \hbox{%
     \vbox{%
       \hrule height 0.5pt 
       \kern0.25ex
       \hbox{%
         \kern-0.1em
         \emath{#1}%
         \kern-0.1em
       }%
     }%
   }%
}

\DeclareRobustCommand{\xbarr}[1]{%
   \hbox{%
     \vbox{%
       \hrule height 0.5pt 
       \kern0.25ex
       \hbox{%
         \kern-0.2em
         \emath{#1}%
         \kern-0.0em
       }%
     }%
   }%
}

\DeclareRobustCommand{\hm}{hydrodynamic memory\xspace}

\DeclareRobustCommand{\bb}{Basset-Boussinesq\xspace}
\DeclareRobustCommand{\bbo}{Basset-Boussinesq-Oseen\xspace}

\DeclareRobustCommand{\ktc}{\emath{k{\!\!\;\scriptstyle{_B}}T_0}\xspace}
\DeclareRobustCommand{\Dx}{\emath{\Delta x}}
\DeclareRobustCommand{\vnet}{\emath{\avg{\overline{v}}}\xspace}

\DeclareRobustCommand{\uosc}{\emath{U_\mathrm{\!\!\;osc}}}
\DeclareRobustCommand{\fosc}{\emath{F_\mathrm{\!\!\;osc}}}
\DeclareRobustCommand{\utilt}{\emath{U_\mathrm{\!\!\;tilt}}}
\DeclareRobustCommand{\vqs}{\emath{v_{qs}}}

\DeclareRobustCommand{\mcE}{\emath{\mathcal{E}}}

\DeclareRobustCommand{\tc}{\emath{t_c}}

\DeclareRobustCommand{\Vc}{\emath{v_c}}
\DeclareRobustCommand{\Mc}{\emath{M_c}}
\DeclareRobustCommand{\Lc}{\emath{L_c}}
\DeclareRobustCommand{\Fc}{\emath{F_c}}

\DeclareRobustCommand{\Fb}{\emath{F_{\scriptstyle_{\!B}}}}
\DeclareRobustCommand{\Fs}{\emath{F_{\scriptstyle_{\!S}}}}
\DeclareRobustCommand{\zst}{\emath{\zeta{\scriptstyle_{\!S}}}}
\DeclareRobustCommand{\taub}{\emath{\tau_{\scriptstyle_{\!B}}}}
\DeclareRobustCommand{\taus}{\emath{\tau_{\scriptstyle_{\!S}}}}
\DeclareRobustCommand{\taunu}{\emath{\tau_\nu}}

\DeclareRobustCommand{\tC}{\emath{\taub}}
\DeclareRobustCommand{\EC}{\emath{\ktc}}
\DeclareRobustCommand{\VC}{\emath{v_{th}}}

\DeclareRobustCommand{\LC}{\emath{\VC\tC}}

\newcommand{\bbp}{BBO particle\xspace}

\newcommand{\bbps}{BBO particles\xspace}

\begin{document}

\preprint{APS/123-QED}
\title{Surmounting potential barriers: hydrodynamic memory hedges against thermal fluctuations in particle transport}

\author{Sean L. Seyler}
\affiliation{Department of Physics, Arizona State University, Tempe, Arizona 85287, USA}
\author{Steve Press\'{e}}
 \email{spresse@asu.edu}
\affiliation{Department of Physics and School of Molecular Sciences, Arizona State University, Tempe, Arizona 85287, USA}

\date{\today}
\begin{abstract}
Recently, trapped-particle experiments have probed the instantaneous velocity of Brownian motion revealing that, at early times, hydrodynamic history forces dominate Stokes damping. In these experiments, nonuniform particle motion is well described by the \bbo (BBO) equation, which captures the unsteady Basset history force at low Reynolds number. Building off of these results, earlier we showed that, at low temperature, \bbps could exploit fluid inertia in order to overcome potential barriers (generically modeled as a tilted washboard) while its Langevin counter-part could not.
Here, we explore the behavior of \bbps at finite temperature. Remarkably, we find that the transport of particles injected into a bumpy potential with sufficiently high barriers can be completely quenched at intermediate temperatures, whereas itinerancy may be possible above and below that temperature window. This effect is present for both Langevin and BBO dynamics, though these occur over drastically different temperature ranges. Furthermore, \hm mitigates these effects by sustaining initial particle momentum, even in the difficult intermediate temperature regime.
\end{abstract}

\maketitle
\raggedbottom

\paragraph{Introduction.} Across many disciplines, growing interest in non-Markovian and general nonequilibrium phenomena has kindled the demand for efficient multiscale models such as generalized Langevin equations (GLEs) \cite{Mori1965-za, Kubo1966-xr, Zwanzig2001-bd}. The \bbo (BBO) equation, which can be expressed as a GLE, has been successfully used to model \hm effects in Brownian particles, capturing the hydrodynamic long-time tail in the velocity autocorrelation that decays with the well-known $t^{-3/2}$ \cite{Alder1970-kg}.
This non-exponential decay is a consequence of viscous coupling between a particle and the vorticity induced by its \emph{unsteady} motion through the ambient fluid. More specifically, the momentum transferred to the fluid remains in the particle's vicinity for a short time $\taunu$ before being carried away through vorticity diffusion \cite{Landau1966-ke, Zwanzig1970-qy, Arminski1979-vi, Maxey1983-gz}. However, if the particle moves substantially in the time $\taunu$, a delayed self-interaction force arises due to viscous feedback from the surrounding fluid. This hydrodynamic force---the \bb history force---acts back on the particle at later times.
As a direct result of this interaction, at finite temperature, thermal fluctuations in the fluid---white Gaussian noise in the viscous stresses \cite{Landau1966-ke, Fox1970-yr}---become time-correlated as they are entrained by the vorticity field about an accelerating particle and manifest as colored (Gaussian) noise
\cite{Mainardi2012-dd, Tothova2016-as, Fodor2015-wq, Franosch2011-ry, Jannasch2011-vf, Kheifets2014-uk}.

Previously, we investigated the \bbps in potentials under time- and space-periodic forcing (generically modeled using potentials such as the tilted washboard) \cite{Seyler2019-rk}.
In that work, we excluded thermal noise and directly compared underdamped Langevin dynamics (i.e., pure Stokes drag and no thermal noise) to BBO dynamics to elucidate the physical mechanism by which \hm---present in the BBO equation---reduces Stokes (i.e., heat) dissipation and, therefore, transport friction. We found that the latent fluid momentum captured by the Basset history force in the BBO equation was sufficient to allow \bbps to reach itinerant states while the particles modeled by underdamped Langevin dynamics remained trapped. Indeed, in a remarkable Letter by \textcite{Goychuk2019-gl}, \hm was seen to have a profound influence on nonlinear diffusion in tilted washboard potentials that extends over much longer timescales than the memoryless case.

Taking \cite{Goychuk2019-gl} and \cite{Seyler2019-rk} together, it is reasonable to imagine that any \bbp that overcomes a fixed barrier height and reaches an itinerant phase at low temperature should continue to do so at higher temperatures. Somewhat counterintuitively, we find that this assertion is not true not only for \bbps but underdamped Langevin dynamics as well, though the onset of trapping occurs at temperatures differing by an order of magnitude.
We elucidate the nontrivial interplay between thermal fluctuations and the force derived from the washboard potential in both the BBO and Langevin equations to explain the origin of trapping over a finite temperature regime from a physical standpoint.
We also describe how \hm fundamentally mitigates the forces that lead to trapping not only promotes itinerancy over a much larger temperature range but also over much longer timescales.

\paragraph{Fluctuating \bbo equation.} Consider a microsphere \cite{Vert2012-wk} of radius $R$, mass $m$, and density $\rho_s$ in an incompressible fluid of density $\rho$ and dynamic viscosity $\eta$, with no-slip boundary conditions. In the limit of low Reynolds and Mach numbers, nonuniform particle motion is described by a BBO equation \cite{Boussinesq1885-tj, Basset1888-ml, Oseen1927-qd}; when such a particle is additionally subjected to thermal noise $\xi(t)$, we will refer to the resulting stochastic equation of motion as the \emph{fluctuating} BBO (FBBO) equation,
\begin{equation}\label{eq:fbbo}
    m_e\dot{v} + \zst v(t) + \zst\sqrt{\frac{\taunu}{\pi}} \!\ii{t_0}{t}
    \!\, \frac{\dot{v}(\tau)}{\sqrt{t-\tau}}d\tau + \xi(t) = f(x),
\end{equation}
where $f(x) = -\pdi{U(x)}{x}$ is a conservative external force.
In the first term, $m_e = \tfrac{2}{3}\pi R^3(2\rho_s + \rho)$ is the \emph{effective mass}, which accounts for the displaced fluid's inertia---an inviscid-unsteady effect; the second term is the quasisteady (Stokes) drag with friction coefficient $\zst = 6\pi\eta R$, while the third is the Basset history force---a consequence of viscous-unsteady flow \cite{Parmar2011-nj}.
By the second fluctuation-dissipation theorem (FDT) \cite{Mori1965-za, Kubo1966-xr}, the history force implies that $\xi(t)$ is colored, containing an anticorrelated Gaussian noise with zero mean and variance \cite{Fodor2015-wq, Tothova2016-as}:
\begin{equation}\label{eq:fdt}
    \avg{\xi(t)\xi(t')} = \kt\zst\left(\delta(t-t') - \sqrt{\frac{\tau_\nu}{4\pi}} \frac{1}{\abs{t-t'}^{3/2}} \right),
\end{equation}
for $t \geq t'$. We refer to \Cref{eq:fbbo} along with the noise statistics in \Cref{eq:fdt} as the \emph{fluctuating BBO} (FBBO) equation.

Two physical timescales appear in \Cref{eq:fbbo}: the particle momentum relaxation time, $\taus = m_e/\zst$ (sometimes \emph{Brownian relaxation time}), and the fluid kinematic time, $\taunu = \rho R^2/\eta$.
Note that in the limit of vanishing fluid inertia ($\rho_s \gg \rho$), vorticity diffusion is effectively instantaneous ($\taunu \ll \taus$). Thus, the history force vanishes and $\xi(t)$ becomes a white noise process. Similarly, \Cref{eq:fbbo} reduces to the conventional Langevin equation (LE) and we obtain underdamped Langevin dynamics (LD).

Following \cite{Seyler2019-rk}, we nondimensionalize \Cref{eq:fbbo} by introducing the following dimensional scales (also cf. \citet{Arminski1979-vi}):
\begin{equation}\label{eq:scaled_variables}
    t_c = \taus \qquad \mcE_c = \ktc \qquad v_c = \VC,
\end{equation}
where $\VC = \sqrt{\ktc/m_e}$ is the thermal speed, yielding the dimensionless time $\tilde{t} = t/\tc$, velocity $\tilde{v} = v/\Vc$, temperature $\tilde{T} = T/T_0$, and so on. Accordingly, length is scaled by $\Lc = \LC$, mass by $\Mc = m_e$, force by $\Fc = \EC/\VC\taus$, and friction by $\zst$.
The dimensionless parameter $\beta$ relates the particle and fluid timescales through $\beta = \tfrac{9\rho}{2\rho_s+\rho} = \taunu/\taus$ so that $\taus = \rho R^2/\beta\eta$ \cite{Arminski1979-vi, Mainardi2012-dd}.
Redefining $\tilde{t}$ as $t$, $\tilde{v}$ as $v$, and so on, we finally have:
\begin{equation}\label{eq:fbbo_nd}
    \dot{v} + v(t) + \sqrt{\frac{\beta}{\pi}}\ii{t_0}{t} \frac{\dot{v}(\tau)}{\sqrt{t-\tau}}d\tau + \xi(t) = -\pdf{U}{x}.
\end{equation}

\begin{figure}[tb]
    \centering
    \vspace{0.25em}
    \includegraphics[width=0.9\columnwidth]{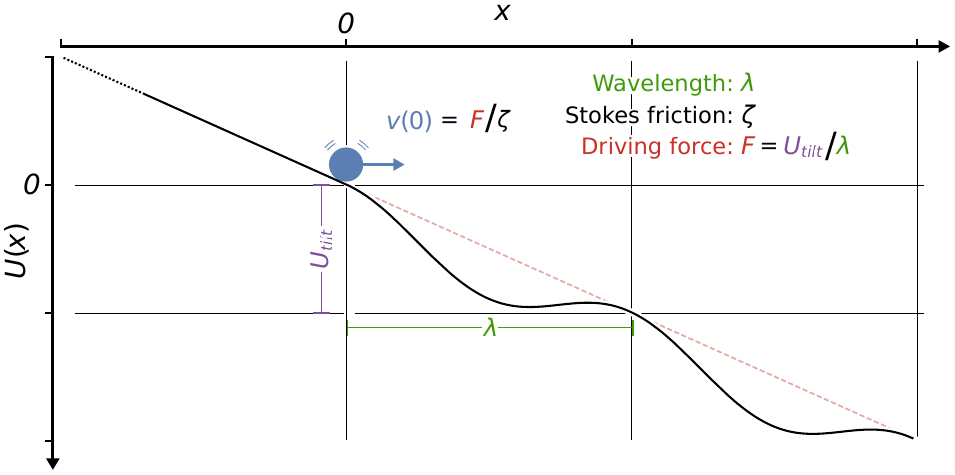}
    \vspace{-1.0em}
    \caption{Schematic of the numerical experiment. For $x(t<0) < 0$, particles move at terminal velocity $v = F/\zst$ under a constant force $F$; at $x(t=0) = 0$, particles enter a tilted washboard potential, $U(x) = \uosc\cos{(2\pi x/\lambda)} - F x$, with wavelength $\lambda$, amplitude $\uosc$, and tilt force $F = 2\pi\utilt/\lambda$.
    }\label{fig:wb_diagram}
    \vspace{-1.5em}
\end{figure}

Here we consider the tilted washboard potential as it is a generic periodic potential with broad physical relevance
\cite{Vollmer1983-or, Jung1984-bs, Ajdari1991-zt, Reguera2000-ub, Gruner1981-sa, Yu2003-le, Coffey2012-mm}
that allows us to explore the concepts of particle trapping (by virtue of this potential's barriers) and itinerancy (by virtue of its periodicity). The tilted washboard potential with wavelength $\lambda$, amplitude $U_0$, and tilt $F$ is given by $U(x) = \uosc\cos{(2\pi x/\lambda)} - F x$. At the critical tilt, $F  = 2\pi \uosc/\lambda$, below which the potential has alternating local minima and barriers; \Cref{fig:wb_diagram} schematically illustrates the washboard potential setup and the initial conditions.

\paragraph{Problem Overview.} We obtain numerical solutions to \Cref{eq:fbbo} using an extended phase space method wherein the memory kernel is approximated as an exponential sum.
This approach---sometimes called a \emph{Prony series} representation or \emph{Markovian embedding}---has been extensively developed and deployed by Goychuk \cite{Goychuk2009-na, Siegle2011-qq, Berg2012-yi, Goychuk2019-gl}
and others \cite{Ferrario1979-mz, Kupferman2004-je, Ottobre2011-rl, Baczewski2013-ts}.
In our previous study \cite{Seyler2019-rk}, we verified the accuracy of this approach in the zero-temperature case by comparison with known analytical solutions \cite{Arminski1979-vi}.
A distinct advantage of the embedding procedure is that the correct thermal noise correlations from the FDT are generated automatically as a weighted sum of Ornstein-Uhlenbeck processes \cite{Baczewski2013-ts}.

We assume neutrally buoyant particles ($\rho_s = \rho$) for FBBO dynamics, which is pertinent to many processes in liquid water. Since $\beta = 3$, the history term is of order unity by \Cref{eq:fbbo_nd} and can only be neglected when $\dot{v} \sim 0$, i.e., when $f$ varies slowly such that the particle travels near terminal (i.e., quasisteady) velocity, $v(t) \sim f[x(t)]/\zst$.

At this point, it bears emphasizing a key difference between the present study and previous studies focusing on nonlinear diffusion enhancement, where it is typical to set $\avg{v(0)} = 0$ and place all particles near a local potential minimum (usually near $x(0) = 0$). In particular, it is typical to study the dynamics around the critical washboard tilt using few select values of washboard amplitude and temperature
\cite{Reguera2000-ub, Reimann2001-fy, Reimann2002-mb, Reimann2008-vk, Evstigneev2008-un,
      Guerin2017-yb, Goychuk2019-gl}.
As such, we fix the tilt, vary both the temperature and washboard amplitude, and quantify transport behavior in terms of the \emph{net velocity},
\begin{equation}
    \avg{\overline{v}(t)} = \frac{1}{t}\ii{0}{t}\avg{v(\tau)}\;\!d\tau = \frac{1}{t}\avg{\Dx(t)},
\end{equation}
where $\Dx(t) = x(t)-x(0)$ is the net displacement. In particular, for transport through a periodic potential, $\avg{\overline{v}(t)}$ can be used as a proxy for the effective transport friction and efficiency \cite{Seyler2019-rk}.
The methodology in the present article connects more straightforwardly to questions of transport efficiency than the viewpoint of enhanced diffusion, as well as revealing nontrivial transport behavior that depends simultaneously on the temperature (noise level), potential amplitude (barriers), and the dynamics (e.g. LD, FBBO).

All particles are injected into the washboard potential at $x(0) = 0$ with the aforementioned initial conditions. One can imagine that for $t < 0$ a particle is in a quasisteady state under the influence of a constant driving force $F$ with no potential oscillations ($\uosc = 0$); at $t = 0$, the particle ``goes off-road'' as it enters the wasboard region ($\uosc > 0$). Because the driving force $F$ is fixed, we anticipate that net velocities will generally decline and possibly drop to zero if barriers are sufficiently high. It is then natural to ask how and to what degree \hm can mitigate the impact of a bumpy landscape and thermal noise on microparticle transport.

Our numerical experiments are germane to the driven transport of microparticles, revealing a remarkable interplay between \hm, thermal fluctuations, and barriers in a bumpy energy landscape. For instance, if we have a particle in motion under the influence of a driving force with hydrodynamic memory, how resistant is the motion to disruption by thermal noise when the amplitude of an oscillatory potential landscape is increased? Does thermal noise help, hurt, or have no effect on a driven particle when it traverses a bumpy potential? How do the underlying dynamics (e.g., overdamped to underdamped, underdamped to \hm) impact the interplay of thermal noise and energy barriers in a spatially periodic potential?

\paragraph{Results.} Simulations were performed with the numerical code used in \cite{Seyler2019-rk} but with noise terms turned on, and leverages the methodology developed by \citetext{Goychuk2019-gl}.
\footnote{We employ the nondimensionalization outlined in the main text. Parameter choices for Markovian embedding closely follow Section III of the SM of \cite{Goychuk2019-gl}, including scaling parameter $b$ and timestep $\Dt = 0.002$. Ensembles averages over 100 particles was satisfactory for present purposes as we calculate $\vnet$, a time-averaged quantity. Numerical integration was performed using a efficient, custom 2nd-order method similar to \cite{Baczewski2013-ts} that we will describe elsewhere; direct comparison to stochastic Heun integration with identical pseudorandom number streams yields very close agreement.}.
The numerical experiments are set up as follows. An ensemble of particles is taken to be in steady state under tilt force $F$, which satisfies the force balance condition, $\avg{\Fs(0) + \Fb(0)} = F$, where $\Fs$ and $\Fb$ are the Stokes and Basset forces, respectively. For underdamped LD ($\Fb \equiv 0$), the velocities are initialized with the mean equal to the terminal velocity $\avg{v(0)} = F/\zst$ and the dispersion given by a thermal distribution $\avg{v^2(0)} = \kt/m_e$.
In the case of \bbps, the auxiliary variables for the Markovian embedding must be initialized appropriately. Force-balance in steady state requires $\avg{\Fb} = 0$ such that the time derivatives of all auxiliary variables vanish simultaneously and $\avg{s_k(0)} = -\gamma_k F/\zst$; auxiliary variables are furthermore given a thermal distribution $\avg{\dot{s}_i(0)\dot{s}_j(0)} = \nu_i\gamma_i \VC\delta_{ij}$, which is required for stationarity of the noise \cite{Siegle2011-qq}.

\begin{figure}[tpb]
    \centering
    \vspace{0.5em}
    \includegraphics[width=1.0\columnwidth]{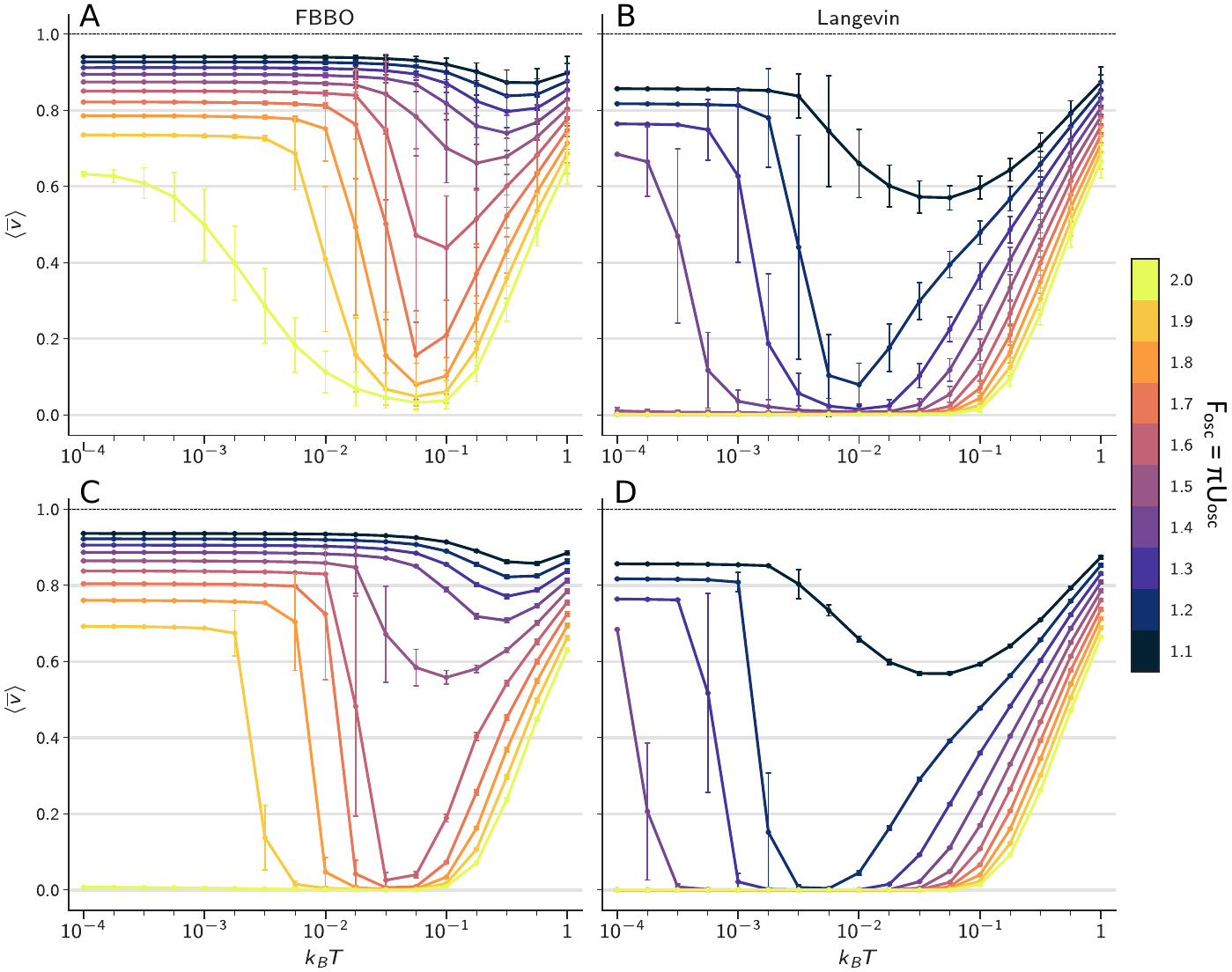}
    \vspace{-2.0em}
    \caption{The sensitivity of the net velocity $\vnet$ of a particle in transport to thermal noise depends on both barrier heights (hue) and the underlying dynamics (FBBO dynamics---left, Langevin dynamics---right). Early on ($t=\num{2d3}$, top row), the dispersion in $\vnet$ is noticeable at higher temperatures. At later times ($t=\num{2d5}$, bottom row), $\vnet$ curves relax as transport is quenched for lower and lower temperatures.
    }\label{fig:vnet_vs_temp}
    \vspace{-1.0em}
\end{figure}

The washboard tilt force is set to $F=1$ and the wavelength to $\lambda = 2$; the washboard force amplitude $\fosc = 2\pi\uosc/\lambda = \pi\uosc$ is varied from 1.1 to 2 in steps of 0.1 so that the washboard is below the critical tilt (i.e., barriers are present) and we vary the temperature between $T = \num{d-4}$ to 1.
The constant force case ($\uosc = 0$) serves as a reference point, where the net velocity $\vnet$ is a maximum $\vqs = F/\zst = 1$, which is simply the quasisteady (i.e., terminal) velocity for both LD and \bbps, by construction. \Cref{fig:vnet_vs_temp} shows, for two different time points, the effect of thermal noise on \vnet over a range of washboard amplitudes. We examine amplitudes larger than the critical value $\uosc = 1.0/\pi$, the value at which barriers vanish (i.e., critical tilt of the washboard).

At $\uosc = 1.1/\pi$, there is already a significant dip in $\vnet$ for LD between $T = \num{8d-3}$ to $\num{8d-2}$, where $\vnet$ is suppressed up to 40\% (\Cref{fig:vnet_vs_temp}B).
At $\uosc = 1.2/\pi$, transport of LD particles is already greatly suppressed between $T = \num{2d-3}$ and $T = \num{2d-2}$ by $t=\num{d3}$ in spite of the fact that at zero temperature, LD particles are itinerant with $\vnet \sim 0.8$. The same qualitative features are evident at $\uosc = 1.6$, where the temperature range of quenched dynamics spans an even wider temperature range, for $\num{4d-4} \lesssim T \lesssim \num{5d-2}$, with the zero-temperature dynamics giving $\vnet \sim 0.68$. Once $V \geq 1.8$, the barrier are sufficiently high that LD particles are no longer itinerant at zero temperature, so transport is possible only at high temperatures $T \gtrsim \num{5d-2}$.

\begin{figure*}[htpb]
    \centering
    \includegraphics[width=\textwidth]{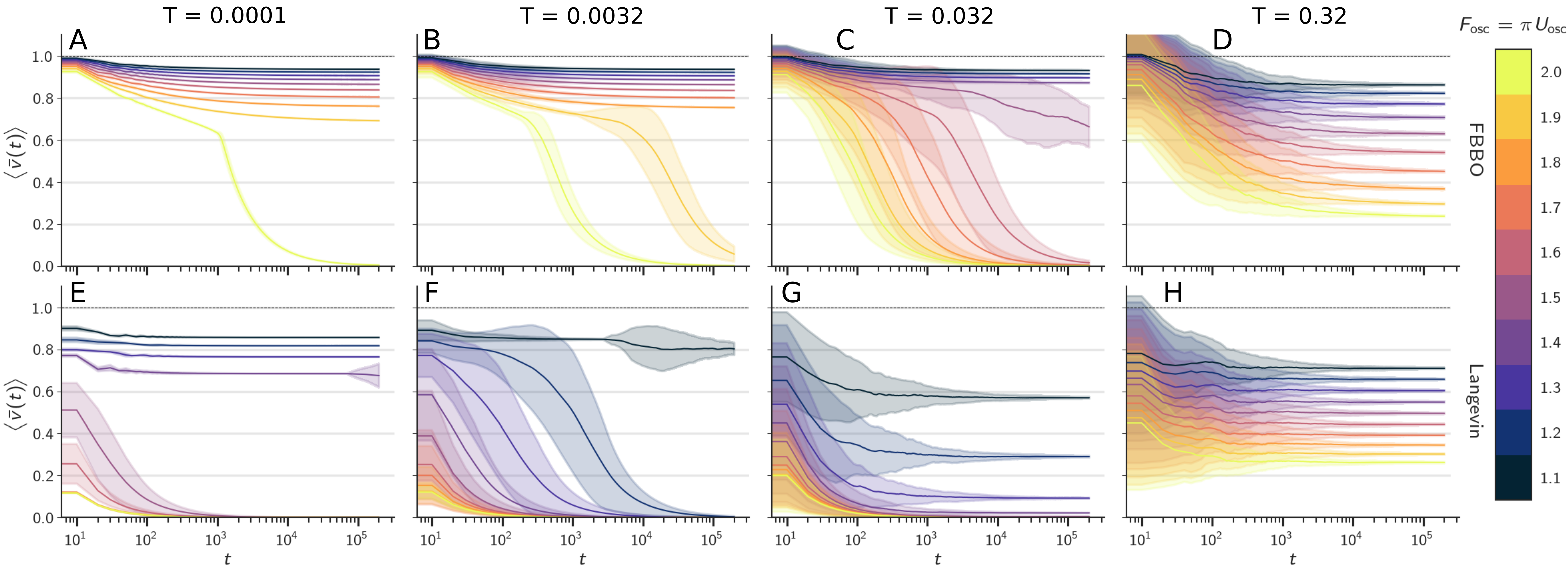}
    \vspace{-2.5em}
    \caption{Net velocity of FBBO (top row) and LD (bottom row) particles as a function of time across four different temperatures (from left to right: $T = \num{d-4}$, $\num{3.2d-3}$, $\num{3.2d-2}$, and $\num{3.2d-1}$.
    }\label{fig:vnet_timeseries}
    \vspace{-1.5em}
\end{figure*}

The presence of \hm only slightly modifies the qualitative features present in \Cref{fig:vnet_vs_temp}. For instance, for both LD and \bbps, $\vnet$ drops sharply and has a large variance for temperatures just below where quenching occurs. The quantitative differences are, however, comparatively dramatic.
For $1.1 < \pi\uosc < 1.4$, $\vnet$ for \bbps gradually decreases across all temperatures for increasing $\uosc$ with a widening dip in the curves just above $T = 0.1$ (\Cref{fig:vnet_vs_temp}A,C).
For barrier heights above $\uosc = 1.5/\pi$, transport is quenched in a narrow temperature range (around $T \sim \num{5d-2}$) that widens dramatically as the amplitude is increased. LD particles, however, exhibit quenched transport for all but the lowest barrier by $t = \num{2d5}$ (\Cref{fig:vnet_vs_temp}D).
Moreover, barriers as low as $\uosc = 1.3/\pi$ suppress LD particle transport even by $t = \num{2d3}$. \Cref{fig:vnet_timeseries} clearly illustrates the comparatively rapid relaxation of $\vnet$ for LD particles at lower temperatures (bottom row, $T < 0.32$)

On the other hand, quenching takes place for \bbps when $\uosc \gtrsim 1.6/\pi$ in a narrow temperature band. \Cref{fig:vnet_vs_temp}C indicates that itinerancy is maintained for barriers as high as $\uosc = 1.5/\pi$, while \Cref{fig:vnet_timeseries} (top row, $T = \num{3.2d-2}$) clearly illustrates the protracted relaxation of $\vnet$ due to \hm.
At the lower end of the temperature range ($T \lesssim \num{d-2}$), $\vnet$ is around 80\% of $\vqs$ for $T \lesssim \num{d-2}$ and, as long as the temperature is low enough ($T < \num{d-3}$), $\vnet$ is still nearly 70\% of $\vqs$.
It is only at $\uosc = 2.0/\pi$ that \bbps can no longer sustain transport at any temperature below $T \sim \num{d-1}$ (\Cref{fig:vnet_timeseries}, top row, $T = \num{d-4}$), at which point zero-temperature transport is not possible, at least with the present initial conditions.


\paragraph{Discussion.}
Intuitively, we expect there to be at least two states of transport, namely itinerant and trapped states \cite{Risken1996-ar}\footnote{What we call \emph{itinerant} and \emph{trapped} states are referred to, respectively, as \emph{running} or \emph{locked} states by \textcite{Risken1996-ar}}. For sufficiently high temperature and/or vanishing potential barriers, we expect an itinerant state regardless of the dynamics (FBBO or LD). For high barriers and low temperatures, we expect particles to become trapped in wells and overall transport to halt. In the zero-temperature limit, it is known that there are situations where \hm enables itinerancy whereas accounting solely for Stokes dissipation predicts trapping \cite{Seyler2019-rk}.

As the temperature is gradually increased, this qualitative difference in transport behavior should persist because we expect the zero-temperature dynamics to smoothly transition into the finite-temperature dynamics. As $T$ is increased further, it becomes increasingly likely that, upon entering the washboard, a particle's motion will be arrested by unfavorable thermal fluctuations, which rob it of the requisite momentum to surmount a barrier and trap in a potential well. If itinerancy is possible for zero-temperature transport, then for small but finite $T$, itinerancy can maintained for a considerable period of time before substantial trapping occurs, though there is a negligible probability of escape once trapped at these temperatures.

Our results clearly demonstrate that \hm not only dramatically delays the onset of trapping but also makes transport more robust to larger barriers.
On the other hand, the probability that a trapped particle will \emph{escape} in a given interval of time (via fortuitous thermal fluctuations) grows with increasing $T$. Concordantly, while \bbps are more resistant to trapping, they are also more difficult to untrap. This is due to both the memory term as well as the anticorrelated hydrodynamic noise, the combination of which extends the time it takes for thermal fluctuations to induce a transition out of a well.
In the high temperature limit ($\kt \gg \uosc$), both FBBO and LD particles are insensitive to the structure of the potential and the overall transport speed is dictated by the tilt $F$.

One might imagine a situation in which the opposing processes of trapping and untrapping are competitive in that they act on roughly the same timescale. Such a situation gives rise the the phenomenon of bistability (i.e., bimodal distribution in the instantaneous velocity), which, broadly speaking, is possible when the washboard potential is below the critical tilt and the Stokes friction is sufficiently small (and inertia significant) \cite{Risken1996-ar}.
Indeed, this phenonmenon was examined in the recent Letter by \textcite{Goychuk2019-gl} for \bbps in tilted washboard potentials. In particular, it was shown that strong diffusion enhancement emerges when particles---with varying strength of hydrodynamic memory---actively transition between itinerant and trapped metastable states.
Having been focused on the physically relevant case of neutrally buoyant (FBBO) particles, our simulations used a relatively large Stokes friction, so little if any bistable behavior was to be expected. It would be fruitful to bridge the present work with \textcite{Goychuk2019-gl} to further elucidate the subtle interplay between hydrodynamic memory and anticorrelated thermal noise with spatially periodic potentials.

Does the temperature window where transport is quenched extend to arbitrarily low $T$ as $t \to\infty$? We do know that when barrier heights are small enough, the zero-temperature dynamics in a tilted washboard will admit particle transport, implying that the quenched transport window cannot contain $T = 0$.
There is a ``corner'' in the $\vnet$ curves where transport drops from finite positive values to essentially zero over a very narrow temperature range. As time goes on, this corner gradually shifts left (toward lower $T$), which reflects the fact that it takes longer to initiate trapping when the thermal noise magnitude is lower. How long does it take for this abrupt switching---from itinerant to trapped dynamics---to take place at a given temperature, and how does the typical quenching time of transport differ between BBO and Langevin dynamics? We reiterate that the present work examined particle transport from the perspective of injecting particles at speed into a bumpy landscape, whereas most studies on nonlinear diffusion consider what is effectively a trapped initial state. Therefore, in light of the long-time influence of \hm observed in the present work and by \cite{Goychuk2019-gl}, how do the differences themselves depend on (the \emph{choice} of) initial conditions?


It is not clear how to disentangle the effect of correlated hydrodynamic noise from the history term---by fluctuation-dissipation, both terms arise simultaneously. However, it seems reasonable to surmise that the anticorrelated noise component has a reduced tendency to induce trapping of a particle already in motion, since this would require a sequence of thermal fluctuations against the motion of a \bbp. On the other hand, the anitcorrelated noise would make it more difficult for a trapped \bbp to escape. In the face of thermal noise and a fluctuating potential, it can be seen that there is a tradeoff between robustness of transport and the ease of untrapping a trapped particle. In the case of general space- and time-dependent driving, one might consider an active transport situation where energy is used to reinitiate transport once a particle is trapped. The question of transport efficiency must then account for the proportion of time a particle spends in the itinerant state and the energy required to move it back to the itinerant state once it is trapped.

Beyond regions of high fluctuations, our work identifies other critical hallmarks of BBO dynamics that suggest experiments to unravel the importance of \hm effects on particle motion.
Comparing FBBO and LD particles after $t = \num{2d5}$, there is over an order of magnitude difference in range of temperatures where transport is quenched. Furthermore, the onset of quenched transport takes place much later when \hm is present for all but the highest temperatures (\Cref{fig:vnet_timeseries}). In principle, such a difference in the relaxation time of $\vnet$ should be experimentally observable.
In particular, experiments in sculpted potentials, such as in Refs
\cite{Curtis2003-xp, Guo2004-rz, Lee2005-mf, Lee2006-ty, Evstigneev2008-un, Siler2010-bb,
    Juniper2012-xl, Jun2014-iw, Gavrilov2016-nm, Kumar2018-ao},
could begin to verify the predictions put forward here.

\clearpage
\newpage


\bibliography{references}


\end{document}